\documentclass[final,authoryear,5p,twocolumn]{elsarticle}
\usepackage{amssymb}
\usepackage{amsmath}
\usepackage{epstopdf}
\usepackage{natbib}
\usepackage{color}

\def \pd{\partial}

\journal{Journal of the International Journal of Solids and Structures}

\begin{document}

\begin{frontmatter}

\title{{\bf {On gradient enriched elasticity theories:\\
A reply to ``Comment on
`On non-singular crack fields
in Helmholtz type enriched elasticity theories' '' \\
and \\important theoretical aspects
}}}
\author[darmstadt]{Markus Lazar\corref{cor1}}
\ead{lazar@fkp.tu-darmstadt.de}
\cortext[cor1]{Corresponding author.
Tel.: +49(0)6151/163686.}
\author[darmstadt]{Eleni Agiasofitou}
\ead{agiasofitou@mechanik.tu-darmstadt.de}
\author[Polyzos]{Demosthenes Polyzos}
\ead{polyzos@mech.upatras.gr}
\address[darmstadt]{Heisenberg Research Group, Department of Physics,
Darmstadt University of Technology,
Hochschulstr.~6, D-64289 Darmstadt, Germany}
\address[Polyzos]{Department of Mechanical and Aeronautical Engineering,
University of Patras, GR-26500 Patras, Greece}

\begin{abstract}
The Comment by Aifantis that criticizes the article
   `On non-singular crack fields
         in Helmholtz type enriched elasticity theories'
        [Lazar, M.,  Polyzos, D., 2014.
         Int. J. Solids Struct.
         doi: 10.1016/j.ijsolstr.2014.01.002]
is refuted by means of clear and straightforward arguments. Important theoretical aspects of gradient enriched elasticity theories which emerge in this work are also discussed.
\end{abstract}

\begin{keyword}
Cracks \sep dislocations \sep fracture mechanics \sep nonlocal elasticity
\sep gradient elasticity\
\end{keyword}

\end{frontmatter}

\section{Introduction}
\label{intro}
A comment by~\citet{A14} [thereafter referred to as (A)]
criticizes some aspects of the article~\citep{LP14} [referred to as (LP)]. In the (LP) article, we discuss and point out the physical or unphysical
meaning and interpretation
of simple non-singular stress fields of cracks of mode I and mode III
published by~\citet{Aifantis09,Aifantis12,Isaksson12} and
~\citet{Isaksson13}. In particular, we investigate if the aforementioned solutions
satisfy the equilibrium conditions, boundary conditions and
compatibility conditions in the framework of nonlocal or gradient elasticity theory.
In the meantime, \citet{KA13} and \citet{I14}
have continued to publish their non-singular stress fields in other
articles in scientific journals.
The aim of the (LP) article was to show that the results of
\citep{Aifantis09,Aifantis12,Isaksson12,Isaksson13}
cannot be the correct solutions of a nonlocal or gradient compatible elastic fracture mechanics problem, since important physical conditions are violated and not to discuss about the physical assumptions of the GRADELA model
made by Aifantis and others. In this work, we reply to the comments of (A) which are relevant to the (LP) article and we clarify important aspects of gradient enriched elasticity theories.

\section{Discussion}
\label{sec2}
We review in the~(LP) article the framework of
Eringen's nonlocal elasticity theory~\citep{E83,E02}. In this context, (A) claimed that ``Eringen never uses the classical stress
$\sigma^0_{ij}$ in a balance law\footnote{Actually, $\sigma^0_{ij,j}=0$ is a conservation law.} $\sigma^0_{ij,j}=0$'', writing additionally that ``Eq.~(LP5) is an artifact of a number of assumptions
and not a fundamental equation of Eringen's theory''. It is easy to show that (A) is mistaken concerning this point due
to the following reasons.
First of all, if the tensor $\sigma^0_{ij}$ is the classical stress tensor,
then it has to fulfill certainly the classical equilibrium condition
$\sigma^0_{ij,j}=0$.
Moreover,
combining Eq.~(LP4) with Eq.~(LP7), one finds
\begin{align}
\label{NE1}
L t_{ij,j}=\sigma^0_{ij,j}=0\,,
\end{align}
which proves that the relation  $\sigma^0_{ij,j}=0$ is true and is part of
Eringen's theory of nonlocal elasticity.
Here,
\begin{align}
\label{L}
L=1-\ell^2\Delta
\end{align}
is the characteristic
differential operator appearing in Eringen's nonlocal elasticity
and in the corresponding gradient elasticity,
$t_{ij}$ is the stress tensor of nonlocal elasticity,
$\ell$ is a characteristic length scale and $\Delta$ denotes the Laplacian. $L$ is usually called as Helmholtz operator in gradient enriched elasticity theories
(see,~\citet{UA95,Vardu,GA99,Peerlings,Paulino03,Gutkin04,LM05,Chan06,Aifantis09,L13}). In general, the Helmholtz equation $\Delta u+c u=0,\, c=\text{constant}$,
 can be found with positive ($c>0$) or negative sign ($c<0$) in the mathematical literature~(e.g., \citet{Smirnov,Roos,Poly,Hassani}); and only sometimes the Helmholtz equation with negative sign, that is when $c<0$, is referred to as the modified Helmholtz equation (e.g.,~\citet{Zauderer}). Moreover, \citet{E83,E02} stated that
in the static case of nonlocal elasticity, the classical equilibrium condition (Eq. (LP5))
is to be satisfied (see, e.g., Eq.~(6.9.32) in \citet{E02}).
Thus, the classical equilibrium condition is part of Eringen's  theory of
nonlocal elasticity in contrast to the claim of (A).
In addition, we note that anisotropic nonlocal elasticity of generalized
Helmholtz type can be found in~\citet{LA11}.
\par
Afterwards, we review in (LP) a simplified gradient elasticity theory, the so-called gradient elasticity of Helmholtz type.
This simplified and straightforward version
of gradient elasticity with only one length scale $\ell$
is a particular case of
Mindlin's theory of gradient elasticity of form~II~\citep{Mindlin64}
and it has been called by~\citet{LMA05}
``gradient elasticity of Helmholtz type''.
The development of gradient elasticity of Helmholtz type~\citep{LM05}
was strongly influenced and stimulated by the field theoretical framework of the
so-called dislocation gauge theory~\citep{L02,L02b,L03,L03b} and
by Feynman's lectures on gravitation~\citep{Feynman} (see also~\citet{LA09,AL10}).
As shown in~\citet{LM05,L13,L14}, gradient elasticity of Helmholtz type
is based on proper variational principles including double stresses
and serves a ``physical''
regularization based on  higher-order partial differential equations.
In \citet{LM05,LM06,L12,L13,L14} gradient elasticity of Helmholtz type has been
successfully applied to calculate non-singular stresses, non-singular
strains and non-singular displacement fields
of straight dislocations and dislocation loops.
It is noted that \citet{LM05} is the first work where such a task was undertaken
within the framework of Mindlin's gradient elasticity with the remarkable outcome that
both the stress and the strain singularities are removed at the dislocation line in contrast to the corresponding solutions of classical elasticity. Moreover, \citet{LM06} corrected the mistaken
gradient elasticity solutions
for the displacement fields of straight screw and edge dislocations
given by~\citet{GA96,GA97,GA99}.
Later, (A)
adopted the correct ``Lazar-Maugin solutions''
in further studies~\citep{Davoudi} (see, e.g.,~\citet{Gutkin06,Davoudi10}).
It should be mentioned that
the non-singular dislocation key-formulas obtained in gradient
elasticity of Helmholtz type~\citep{L13,L14}
have been recently implemented in 3D discrete
dislocation dynamics~\citep{Po14}. The $J$-integral was given by~\citet{LK07} for incompatible gradient elasticity of Mindlin and Helmholtz type and by~\citet{AL09} for Mindlin's gradient elasticity of grade three. In addition, \citet{L14} shows that
the framework of gradient elasticity of Helmholtz type is in
agreement with the framework of the magnetostatic gradient theory,
which is the magnetostatic part of the so-called Bopp-Podolsky theory
\citep{Bopp,Podolsky}
and how gradient theories are used in physics.
However, it has to be mentioned that \citet{Mindlin64}
introduced his theory of gradient elasticity
without giving any credit to Bopp and Podolsky.
\citet{Guenther76} was
the first to speak of a mechanical model of the Bopp-Podolsky potential for
defects in elasticity.
\par
On the other hand, about three decades after~\citet{Mindlin64},
\citet{A92,AA92,RA93,GA96,GA97} introduced a simple gradient elasticity
model called GRADELA.
In fact, \citet{A92} has intuitively postulated a gradient
modification of Hooke's law.
However, the framework of \citet{A92,AA92,GA96,GA97} lacks
double stresses and, consequently, the stress tensor remains singular.
Their approach postulates a sophisticated Hooke's law including a
Laplacian of the elastic strain tensor.
In a refined version of their model, \citet{GA99} found
non-singular stress and non-singular strain fields
using a more sophisticated Hooke's law including
a Laplacian of the (Cauchy) stress tensor.
Moreover, in contrast to Mindlin's theory of gradient elasticity~\citep{Mindlin64}
(see also~\citet{Jaunz}),
the framework of \citet{A92,AA92} and
\citet{GA96,GA97,GA99} is not based on proper variational
considerations (e.g. to obtain pertinent boundary conditions).
Thus, GRADELA is based on ad-hoc postulations and not on a proper
theoretical framework as basis of a theory
(see also~\citet{Fafalis} for an interesting discussion).
It becomes evident
that gradient elasticity of Helmholtz type
is different from
GRADELA due to the aforementioned reasons
and that the authors of the (LP) article have not adopted GRADELA.
\par
Now, we sketch the basics of Mindlin's theory of gradient elasticity of form~II~\citep{Mindlin64}
and we give its relation to the theory of gradient elasticity of Helmholtz type.
In Mindlin's theory of gradient elasticity of form~II, the strain energy density has the
general form
\begin{align}
\label{W}
W=W(e_{ij}, e_{ij,k})\,,
\end{align}
where $e_{ij}$ denotes the elastic strain tensor of gradient elasticity.
The stresses of gradient elasticity are defined as (see, also~\citet{Eshel,Eshel73})
\begin{align}
\label{CR1}
\tau_{ij}&\equiv \frac{\pd W}{\pd e_{ij}}=\tau_{ji}&&\text{Cauchy stress tensor}\,,\\
\label{CR2}
\tau_{ijk}&\equiv \frac{\pd W}{\pd e_{ij,k}}=\tau_{jik}&&\text{double stress tensor}\,.
\end{align}
Thus, $\tau_{ij}$ is the Cauchy stress tensor of gradient elasticity which usually
reads
\begin{align}
\label{HL}
\tau_{ij}=C_{ijkl}e_{kl}\,,
\end{align}
where $C_{ijkl}$ is the tensor of the elastic constants.
In the (LP) article, we use the notation Cauchy-like stress tensor for
$\tau_{ij}$, which means Cauchy stress tensor of gradient elasticity.
The equilibrium condition for the stresses takes the form~\citep{Mindlin64}
\begin{align}
\label{EC}
\tau_{ij,j}-\tau_{ijk,kj}=0\,.
\end{align}
The relation of Mindlin's theory of gradient elasticity of form~II to gradient elasticity of Helmholtz type is given by the particular
form of the double stress tensor~(\ref{CR2})
\begin{align}
\label{CR2-HE}
\tau_{ijk}=\ell^2\tau_{ij,k}\,.
\end{align}
The double stress tensor~(\ref{CR2-HE}) is nothing but the gradient of the Cauchy stress tensor
of gradient elasticity multiplied by the pre-factor $\ell^2$ in order to have the correct dimension
(see, e.g.,~\citet{LM05,Polyzos}). Eq.~(\ref{CR2-HE}) can be easily obtained
by Eq.~(\ref{CR2}) using the strain energy density
of Mindlin's theory of Form II (see Eq.~(11.3) in~\citet{Mindlin64}) with an appropriate simplification of the five gradient coefficients (see, e.g.,~\citet{LM05}). The equilibrium condition~(\ref{EC}) with the help of the relation~(\ref{CR2-HE}) can be reduced to
\begin{align}
\label{EC-HE}
L\tau_{ij,j}=0\,,
\end{align}
where $L$ is the Helmholtz operator given in Eq.~(\ref{L}). Eq.~(\ref{EC-HE}) is the equilibrium condition in gradient elasticity of Helmholtz type. Furthermore, substituting Eq.~(\ref{HL}) and a compatible elastic strain tensor
$e_{ij}=1/2 (u_{i,j}+u_{j,i})$ into Eq.~(\ref{EC-HE}), one can find a
homogenous Helmholtz-Navier equation for the displacement vector $u_k$
\begin{align}
\label{HN-u}
L C_{ijkl} u_{k,lj}=0\,,
\end{align}
which is an elliptic partial differential equation of fourth-order.
\par
Using an operator split (see, also~\citet{Vekua,L14,RA93}),
one can decompose the partial differential equation of third-order~(\ref{EC-HE}) into
a system of partial differential equations consisting of a partial
differential equation of first-order and a partial differential equation of
second-order, as follows
\begin{align}
\label{EC-0}
&\sigma^0_{ij,j}=0\,,\\
\label{HE-g}
&L\tau_{ij}=\sigma^0_{ij}\,.
\end{align}
Eq.~(\ref{HE-g}) is an inhomogeneous Helmholtz equation where the inhomogeneous part
is given by $\sigma_{ij}^0$, which has to fulfill Eq.~(\ref{EC-0}).
Since $\sigma^0_{ij}$ satisfies the classical equilibrium condition~(\ref{EC-0})
and does not
depend on the length scale~$\ell$, it may be identified with the
Cauchy stress tensor of classical elasticity theory (see also~\citet{L14}).
If one substitutes Eq.~(\ref{HL}) into Eq.~(\ref{HE-g}), then one obtains
\begin{align}
\label{HE-HL}
L C_{ijkl} e_{kl}=\sigma^0_{ij} \,,
\end{align}
which is a partial differential equation for the elastic strain tensor of gradient
elasticity of Helmholtz type.
Eq.~(\ref{HE-HL}) should not be confused with a kind of modified Hooke's law;
what is done in the ``basic postulate'' of GRADELA~\citep{A92,AA92,GA96,GA97}.
\par
One of the main results of the (LP) article (what (A) says as the first major criticism of the (LP)) is that: ``Eqs. (30) and (31) are not solution for a mode III crack in nonlocal elasticity since they do not satisfy the equilibrium condition (4)". This is commented by (A) in the framework of gradient elasticity and not in the framework of nonlocal elasticity causing great confusion to the reader. Actually, the comment takes place in the wrong framework. We would like to mention again that the non-singular stresses of mode~III,
Eqs. (LP30) and (LP31), as well as of mode~I, Eqs.~(LP47)--(LP49),
do not satisfy the equilibrium condition~(LP4) of nonlocal elasticity.
Thus, the mentioned nonlocal stresses cannot be considered as
physical solutions in nonlocal elasticity, since they produce
forces at the corresponding crack tips as shown in Eq.~(LP35)  for mode III and in Eqs.
(LP51) and (LP52)  for mode I.
\par
Another important result of the (LP) article (what (A) says as the second major criticism of the (LP)) is that the non-singular stresses and the corresponding
non-singular strains of mode~III and mode~I do not fulfill
the compatibility conditions in the framework of gradient elasticity and they give rise to non-vanishing
dislocation density contributions at the crack tips. In particular, (A) comments the fact that the strains (33) and (34) exhibited in (LP)
   article (originally published in~\citet{Aifantis09}) are
   not compatible by writing that the strains (46) in~\citet{LM05} are also
   not compatible; stating moreover erroneously that there is a contradiction
   between (LP) article and~\citet{LM05} concerning this point. It should
 be here clarified that there is no contradiction,
   since there is a great difference between the two
   problems. \citet{LM05} investigate a problem of dislocations in the
   framework of incompatible gradient elasticity theory, where the strains are incompatible. Whereas, \citet{Aifantis09,Aifantis12,Isaksson12}
   and~\citet{Isaksson13} investigate a problem of cracks and starting from
   classical compatible elastic distortions, they obtain incompatible elastic
   distortions due to gradient elasticity.
   In other words, gradient elasticity creates incompatibilities (dislocations)
   in such a problem of cracks as an artifact.
\par
Furthermore, it is misleading to introduce ``microstress'' and
``macrostress'' in gradient elasticity
as done in (A). Concepts of ``microstress'' and
``macrostress'' make sense in microcontinuum theories like micromorphic elasticity, where
microstrain is defined and contained (see, e.g.,~\citet{E99}).
\par
Moreover, we show that the fact that the stresses (\ref{HL}) and (\ref{CR2-HE}) are
divergence-free, that is  $\tau_{ij,j}=0$ and $\tau_{ijk,kj}=0$, is an outcome of the
theory and it is not restrictive as it is claimed by (A). These properties of the stresses were first pointed out by~\citet{LM05}. Let us discuss first some properties of the solution of the
inhomogeneous Helmholtz equation~(\ref{HE-g}) or (LP19) appearing in gradient elasticity
of Helmholtz type.
Using the theory of partial differential equations~(see, e.g.,~\citet{Wl}),
we obtain the representation of the stress tensor $\tau_{ij}$
as convolution of the Green function $G$ of the Helmholtz equation,
$LG=\delta$, with the classical singular stress tensor $\sigma_{ij}^0$
\begin{align}
\label{T-con}
\tau_{ij}=G*\sigma_{ij}^0\,,
\end{align}
which is the (particular) solution of the
inhomogeneous Helmholtz equation~(\ref{HE-g}). $\delta$ is the Dirac delta function
and $*$ denotes the spatial convolution.
As discussed in~\citet{L14}, the tensor $\tau_{ij}$ has the physical meaning
of the Cauchy stress tensor of gradient elasticity; an interpretation which is
in full agreement with the interpretation used by \citet{Eshel,Eshel73,G1,G2,G3}.
By applying the Helmholtz operator~$L$ to both sides of
Eq.~(\ref{T-con}), the result reads (see, e.g.,~\cite{Kanwal})
\begin{align}
\label{L-P}
L\tau_{ij}=L(G*\sigma_{ij}^0)=(L G)*\sigma_{ij}^0=
\delta*\sigma_{ij}^0=\sigma_{ij}^0\,,
\end{align}
which proves that Eq.~(\ref{T-con}) solves Eq.~(\ref{HE-g}).
Such solution is unique in the class of generalized functions.
If we use Eq.~(\ref{T-con}), the property of the differentiation
of a convolution and that the operation of convolution is
commutative~\citep{Wl,Kanwal}, then we find that the divergence of the
Cauchy stress tensor of gradient elasticity is zero
\begin{align}
\label{DivT}
 \tau_{ij,j}=(G*\sigma_{ij}^0)_{,j}=G*(\sigma_{ij,j}^0)=0\,,
\end{align}
since $\sigma_{ij,j}^0=0$ (see Eq.~(\ref{EC-0})). Hence, $\tau_{ij,j}=0$. Next, we show that $\tau_{ijk,kj}=0$. In order to differentiate a convolution,
it suffices to differentiate any one of the factors~\citep{Kanwal}.
Therefore,
if the convolution~(\ref{T-con}) exists, then
the Cauchy stress tensor of gradient elasticity is self-equilibrated.
If one uses a so-called ``Bifield'' ansatz for the stress \citep{L14}
\begin{align}
\label{BF}
\tau_{ij}=\sigma_{ij}^0+\sigma_{ij}^1\,,
\end{align}
where the tensor $\sigma_{ij}^0$ is the classical stress tensor
and the tensor $\sigma_{ij}^1$, which is the gradient part of the stress,
corresponds to the relative stress tensor (see~\citet{L14}),
then  $\sigma^1_{ij,j}=0$ follows from Eq.~(\ref{DivT}).
Moreover, the double stress tensor~(\ref{CR2-HE}) (see Eq.~(LP11)) is equilibrated by
the relative stress tensor $\sigma_{ij}^1$ (see also~\citet{VS95})
\begin{align}
\sigma_{ij}^1=\tau_{ijk,k}=\ell^2\Delta \tau_{ij}
\,.
\end{align}
It is easy to see that the following relation  holds
\begin{align}
\sigma_{ij,j}^1=\tau_{ijk,kj}=\ell^2\Delta \tau_{ij,j}=0\,,
\end{align}
but $\tau_{ijk,k}\neq 0$. Therefore, we have shown that the stresses $\tau_{ij}$ and $\tau_{ijk}$ are
divergence-free, $\tau_{ij,j}=0$ and $\tau_{ijk,kj}=0$, and this result arises naturally from the theory.
Note that the non-singular gradient-enriched stresses of dislocations and disclinations
given by \citet{GA99,GA00} are also divergence-free. Additionally, using Eq.~(\ref{BF}), the boundary conditions of gradient elasticity
in terms of $\sigma_{ij}^0$ and $\sigma_{ij}^1$ can be
found in~\citet{Ara11} and \citet{L14}.
\par
The $r^{-3/2}$ behavior of the ``total'' stresses
at the crack tip in the framework of strain gradient elasticity 
is known since 2000; confirmed theoretically
(e.g, \citet{AG,FCP,G03,GG09,GG10,SHH,SG})
and numerically (e.g.,~\citet{AZ,Karlis,MTT,PZ}).
Most of these papers should be known to Aifantis for many years
(see, e.g., in~\citet{Aifantis09,Aifantis12}).
 An appropriate treatment of a crack
    problem in the framework of strain gradient elasticity is to solve the
    homogeneous Helmholtz-Navier equation for the displacement vector
    (Eq.~(\ref{HN-u})), since a crack problem is a compatible elasticity
    problem. Moreover, Eq.~(\ref{HN-u}) has to be accompanied by the
    corresponding boundary conditions of gradient elasticity, the so-called
    Mindlin boundary conditions (see Eqs.~(LP10)) which are stemming from variational principles, and by the appropriate regularity conditions. One can see clearly this procedure in~\cite{G03} for a mode III crack problem. Then, the ``total'' stress becomes singular with $r^{-3/2}$~singularity at the crack tip, as it was predicted in the aforementioned theoretical and
numerical works. In addition, \citet{GG09} solved the homogeneous Helmholtz-Navier equation~(\ref{HN-u})
of a compatible elasticity problem for mode I and mode II cracks
together with the appropriate boundary and regularity conditions without using the ``Ru-Aifantis theorem''. They found that even if the ``total'' stress is hypersingular, the ``monopolar'' stress tensor which
is directly connected with the compatible
elastic strain tensor (and corresponds
to the Cauchy stress tensor of gradient elasticity in our notation) is bounded
in the vicinity of the crack tip. 
Moreover, the strain field is also bounded at the crack tip region.

Later, (A) writes about his students and very personal views.
In the rest of his comment (A) discusses some unpublished results
which are related to forthcoming papers and not to the
results of (A) criticized in the (LP) article.
Therefore, there is no need to discuss these items.

\section{Conclusion}
The criticisms of (A) against results in (LP) are refuted by means
of clear and straightforward arguments from the mathematical and physical
point of view. The present discussion reveals also that consistency is
of major significance, since the two comments of (A) against the two major
results of the (LP) article take place in the wrong framework; confusing and
mixing nonlocal elasticity with gradient elasticity and compatible elasticity
with incompatible elasticity. 
A problem of cracks has a different treatment than a problem of dislocations.
\par
Therefore, we conclude again that the non-singular stresses given by
  \citet{Isaksson12} cannot be considered as physical crack solutions in
  nonlocal elasticity and the non-singular strains given by~\citet{Aifantis09, Aifantis12} and~\citet{Isaksson13} cannot be considered
as physical crack solutions
in strain gradient elasticity. In nonlocal elasticity the equilibrium condition is not satisfied,
and in gradient elasticity the compatibility condition is not
fulfilled. In addition, the non-singular solutions of~\citet{Aifantis12,Isaksson12,Isaksson13}
for a mode~I crack do not satisfy the zero stress boundary condition at the crack faces.

\section*{Acknowledgements}
The authors wish to express their thanks to Prof. H.~Georgiadis
for helpful comments concerning the (LP) article.
M.L. thanks Professors H.~Gao, S.~Li and T.~Michelitsch
for useful remarks on the (LP) article.
D.P. thanks Prof. D.~Beskos for useful discussions.
M.L. and E.A. gratefully acknowledge the grants from the
Deutsche Forschungsgemeinschaft (Grant Nos. La1974/2-2, La1974/3-1).

\end{document}